# Combination of informational storage and logical processing based on an all-oxide asymmetric multiferroic tunnel junction

Q. Liu,[1] J. Miao,[1,a)] Z. D. Xu,[2] P. F. Liu,[1] Q. H. Zhang,[3] L. Gu,[3] K. K. Meng,[1] X. G. Xu,[1] J. K. Chen,[1] Y. Wu,[1] and Y. Jiang[1,b)]

[1] Beijing Advanced Innovation Center for Materials Genome Engineering, School of Materials Science and Engineering, University of Science and Technology Beijing, Beijing 100083, China

[2] Department of Physics, Southern University of Science and Technology, Shenzhen, Guangdong 518055, China

[3] Beijing National Laboratory for Condensed Matter Physics, Institute of Physics, Chinese Academy of Sciences, Beijing 100190, China

Multiferroic tunnel junctions (MFTJs) have already been proved to be promising candidates for application in spintronics devices. The coupling between tunnel magnetoresistance (TMR) and tunnel electroresistance (TER) in MFTJs can provide four distinct resistive states in a single memory cell. Here we show that in an all-oxide asymmetric MFTJ of $La_{0.7}Sr_{0.3}MnO_3$ /$PbZr_{0.2}Ti_{0.8}O_3$ /$La_{0.7}Te_{0.3}MnO_3$ (LSMO/PZT/LTMO) with p-type and n-type electrodes, the intrinsic rectification is observed and can be modified by the ferroelectric polarization of PZT. Owing to the combined TMR, TER and diode effects, two different groups of four resistive states under opposite reading biases are performed. With two parallel asymmetric junctions and the appropriate series resistance, the coexistence of logic units and quaternary memory cells can be realized in the same array devices. The asymmetric MFTJ structure enables more possibilities for designing next generation of multi-states memory and logical devices with higher storage density, lower energy consumption and significantly increased integration level.

---

a) Electronic mail: j.miao@ustb.edu.cn

b) Electronic mail: yjiang@ustb.edu.cn

The rapid development of spintronics has brought about the revolution of modern storage technology.[1,2] In order to break through the limit of dimensions and reduce the power dissipation, various emerging materials have been studied for possible applications in information processing and storage.[3-5] Among them, multiferroic tunneling junction (MFTJ), which is composed of two ferromagnetic electrodes separated by a nanometer-thick ferroelectric tunnel barrier, has attracted much attentions.[6-8] In MFTJ, tunnel magnetoresistance (TMR) is spin-dependent and tunnel electroresistance (TER) is modulated by the polarization of ferroelectric barrier.[9-11] The simultaneous existence of TMR and TER may enable novel types of spin control in high-density and low-power-consumption non-volatile memories.[12,13] For example, it has been proposed to store four states in a single MFTJ memory cell, therefore quaternary information is encoded by both ferromagnetic and ferroelectric orders and the non-destructive information can be read by resistance measurements,[6,7] which shall largely increase the information storage density.[9,14,15]

Meanwhile, the interfacial coupling between ferroelectric barrier and ferromagnetic electrodes in MFTJ plays a critical role in TMR and TER.[15-18] Hence, a lot of studies have been concentrated in the interactions in the vicinity of interface such as lattice strain, screening charge, orbital hybridization and so on.[19-25] All the reported manganese oxide-based MFTJs are focused on the symmetric systems in which the bottom and top electrodes use the same kind of hole-doped oxides (such as LSMO).[9,10,13,14,16,18] The symmetric electrodes usually bring symmetric interfacial formation and electrical behaviors.[16,18]

In a doped manganate oxide $RE_{1-x}AE_xMnO_3$ (RE: rare-earth ion, AE: alkaline-earth ion) with perovskite structure, the A-site doping type and level play a decisive role in its electrical, magnetic and structural properties.[26-29] For example, the parent compound $LaMnO_3$ doped with divalent or quadrivalent cations exhibit a rich variety of phenomena including different electrical transport and magnetic properties associated with different carrier types.[26,27,30-32] We expect both the opposite carriers from differently doped electrodes and the switchable ferroelectric barrier can influence the coupling between TMR and TER effects. In the present study, half-metallic oxide $La_{0.7}Sr_{0.3}MnO_3$ (LSMO) with a nominal spin polarization of ~100% serves as a hole-doped p-type electrode. The electron-doped oxide $La_{0.7}Te_{0.3}MnO_3$ (LTMO) serves as a n-type ferromagnetic electrode below its transition temperature (about 200 K).[33] An ultrathin ferroelectric barrier $PbZr_{0.2}Ti_{0.8}O_3$ (PZT) is inserted between the two differently doped electrodes and the fabricated structure of LSMO/PZT/LTMO is an asymmetric MFTJ. The asymmetric structure shows remarkable multiple resistive states as a MFTJ device as well as a remarkable diode effect as a special p-i-n junction with an ultrathin insulator layer. The TMR, TER and diode effects are coexisting and coupling in the asymmetric MFTJ which can influence the tunneling effect and conducting behaviors. This research will be of great significance for realizing novel devices combining high-density memories and diode effect.

**Results**

**Multiferroic tunnel junction devices.** Epitaxial multiferroic multilayers of LSMO (30 nm)/PZT (3.6 nm)/LTMO (15 nm) were deposited by pulsed laser deposition (PLD) on (100)-oriented $SrTiO_3$ (STO) single crystal substrates with a single $TiO_2$-terminated surface. More details of the growth are provided in the 'Methods' section. As shown in Supplementary Figure 1a and 1b, the surface of the substrates is atomically flat. After the deposition of the multilayers, MFTJs in square shape (20 μm ×20 μm) were patterned by electron beam lithography (EBL), ion-milling etching and lift-off process. A polycrystalline $Al_2O_3$ film deposited by RF sputtering was used to isolate bottom LSMO electrode from top Pt electrode.

**Ferroelectric and structural characterization.** As shown in Figure 1a, the step-terrace surface morphology with a step height of 0.4 nm measured by atomic force microscopy (AFM) indicates a layer-by-layer growth. The line tracing of surface height is shown in Supplementary Figure 1c and 1d. Piezoresponse force microscopy (PFM) was used to characterize ferroelectric properties of the ultrathin PZT films. The PFM measurements were performed at room temperature using a conductive tip to scan the surface of the PZT film with a LSMO electrode layer buffered on a STO substrate. The PZT layer was poled upwards and downwards with the writing bias of ±5 V, respectively. The writing bias was selected to ensure complete ferroelectric reversal based on the measurements in Supplementary Figure 2. As in Figure 1b, the ferroelectric phases with clear domain boundaries justify excellent and reversible ferroelectricity of the PZT tunnel barrier. The PFM amplitude and phase loops in Figure 1c demonstrate two antiparallel polarizations with 180° phase contrast controlled by the written bias. Besides, it can be inferred that the spontaneous polarization of PZT is upwards (away from the LSMO bottom electrode) due to appropriate electrostatic boundary conditions.[24,34,35] The asymmetry of PFM amplitude and phase loops should be caused by the spontaneous polarization of PZT. The reciprocal space mapping (RSM) around the asymmetric (103) reflections is shown in Figure 1d for the MFTJ multilayers. The diffraction spots from the corresponding layers and STO substrate show almost identical in-plane scattering vectors $Q_x$, which indicates a well in-plane matching between the thin films and the substrate, and no obvious strain relaxation.[36,37] The larger values of out-of-plane scattering vectors $Q_z$ for LSMO and LTMO are associated with the compression of the out of plane lattice constant $c$ due to the epitaxial strain from the STO substrate.[37] Compared with the perfect cubic unit cell ($c/a$=1), the unit cells of the LSMO and LTMO films deposited on the STO substrate are both out of plane compressed ($c/a$=0.986 for LSMO and $c/a$=0.998 for LTMO). The diffraction spots from the top LTMO layer exhibit slight deviations from the uniform in-plane scattering vectors, which is due to the imperfect crystallization at the deposition temperature of 725 ºC. It should be noticed that no obvious diffraction spot of the ultrathin PZT layer can be detected.

**Interfacial chemical mapping.** To further elucidate the structural features of the MFTJ multilayer, high-resolution transmission electron microscope (HRTEM) was carried out. Figure 2a shows an aberration-corrected HRTEM high-angle annular dark-field (HRTEM-HAADF) image obtained from the MFTJ multilayer. The clear lattice and interface in the absence of obvious intermixing suggests the single crystalline and fully epitaxial growth of all the layers. 9 unit cells confirm the ~3.6 nm thickness of the PZT barrier. The displacement of Ti ions in PZT unit cells was upward along the [001] direction, as shown in the inset of Figure 2a, suggesting an upward spontaneous polarization of the PZT barrier in the initial state of the MFTJ multilayer in agreement with the PFM image in Figure 1 (b). The intensity profile demonstrates the atomically sharp interfaces of LSMO/PZT and PZT/LTMO with the Ti-O-La/Sr (or La/Te) terminations, as shown in Supplementary Figure 3a and 3b). The elemental profiles of Ti and Mn obtained from electron energy loss spectroscopy (EELS) presented in Figure 2b demonstrate the atomic arrangement of manganate and titanite and clearly show that the intermixing only take place at the Mn-O-Ti interfaces (i.e. LSMO/PZT and PZT/LTMO).

EELS line scans for the Mn $L_{2,3}$ edges from the manganate layers in the vicinity of interfaces were collected and the mapping under different layer positions normal to the interfaces are shown in Figure 2c and 2d. It is known that the changes in the Mn $L_{2,3}$ edge spectra are associated with the excitations at the spin-orbit split $2p_{3/2}$ and $2p_{1/2}$ levels to available states in the 3d band.[38,39] The energy shift, $L_{2,3}$ edge separation ($\Delta E_{\text{Mn } L2,3}$) and intensity ratios ($I_{L3}/I_{L2}$) of Mn $L_{2,3}$ edge can reflect the chemical changes of Mn ion.[18,24,25] Comparing the EELS mapping between top LTMO and bottom LSMO layers, there are slight shifts to lower energy in the vicinity of the PZT layer, indicating local chemical state changes.[18,25]

The peak separation ($\Delta E_{\text{Mn } L2,3}$) and the peak intensity ratio $L_3/L_2$ corresponding to the Mn signal profiles for the top LTMO and bottom LSMO layers in the vicinity of the PZT interfaces are presented in Figure 3a and 3b, respectively. The $\Delta E_{\text{Mn } L2,3}$ and $I_{L3}/I_{L2}$ ratio from the top LTMO and bottom LSMO layers towards the PZT interfaces show opposite variation trends in 2~3 unit cell regions. The $L_3/L_2$ ratio calculated using the Pearson method[40] is directly corresponding to Mn valence, and the chemical valence of Mn ions can be estimated using a standard curve representing chemical valence vs $L_3/L_2$ obtained from previous works[41] (as shown in Supplementary Figure 4). The mappings of Mn valence in the vicinity of the PZT interfaces for the top

LTMO and bottom LSMO are shown in Figure 3c and 3d, respectively. The Mn valence of top LTMO layer retains its bulk value of about +2.8 and shows a drop of ~0.2 over two unit cell region above the PZT/LTMO interface. In contrast, the Mn valence of bottom LSMO layer retains its bulk value about +3.0 and shows a large rise of about 0.3 over 1 nanometer below the LSMO/PZT interface.[42] Synchrotron radiation X-ray absorption spectra (XAS) of Ti $L_{2,3}$ edge from the whole PZT layer and the top interface of PZT (the schematic of the measurement is shown in Supplementary Figure 5) are presented in Supplementary Figure 6a. The peaks of B and D denoting $e_g$ orbital from the top interface of PZT exhibit a visible shift towards high energy which indicates a higher chemical state of Ti ions in the vicinity of the PZT/LTMO interface. The opposite chemical state changes of Mn and Ti ions at the top interface verify the interfacial charge transfer.[43] Supplementary Figure 6b schematically shows the motion of charged oxygen vacancy driven by the depolarization field of PZT in the vicinity of the PZT interfaces, which changes the chemical states of Mn ions over two unit cell region. The slight deviation in valence values is basically consistent with the EELS mapping.

**Electrical characterization.** A schematic of the fabricated LSMO/PZT/LTMO (or SPT, for short) MFTJ device and electrical measurements is illustrated in Figure 4a. The magnetic measurements in the direction parallel to film plane were carried out at 60 K before the device fabrication, as shown in Supplementary Figure 7. The structure exhibits a wasp-waist shaped hysteresis loop consisting of two step magnetization reversal, which provides a prerequisite to realize TMR effect.[44] The electrical transport properties of the SPT MFTJ influenced by the ferroelectric polarization of PZT were investigated in the absence of magnetic field at 60 K. In Figure 4a, the I-V curves of the SPT sample show remarkable asymmetry under positive and negative bias. As a special oxide p-i-n junction, our device shows remarkable rectifying characteristics which have not been reported before in symmetric MFTJ systems. The rectification (defined as the ratio of $I_{+0.2V}/I_{-0.2V}$) is 6.52 for the upward polarized PZT and 8.14 for the downward polarized one. The different rectified transport properties influenced by the polarization of PZT are due to the enhanced or weakened space charge region and built-in field.[45] The asymmetric tunneling conductance (*dI/dV*) calculated from the I-V curves show different levels for the opposite polarization fields of PZT (see Supplementary Figure 8d) may be induced by the modified interfacial barrier height.

To further study the dependence of the TMR effect on the ferroelectric polarization of PZT, the resistance curves as a function of magnetic field (R−H) were measured by applying an in-plane magnetic field parallel to the substrate [110] direction and using a pulse bias of -100 mV/100 ms at 60 K, as shown Figure 4b. Before that, the ferroelectric polarization of the PZT barrier layer was pre-written with ± 5 V pulse voltages respectively. Typical four distinct resistive states are realized with the different ferroelectric polarizations. We define the TER and TMR ratios by Equation (1) and (2) respectively.

$$\text{TER} = \frac{R_{down} - R_{up}}{R_{up}} \times 100\% \qquad , \qquad (1)$$

where $R_{up}$ and $R_{down}$ denote the resistances for the upward and downward ferroelectric polarizations of PZT, respectively.

$$\text{TMR} = \frac{R_{AP} - R_P}{R_P} \times 100\% \qquad , \qquad (2)$$

where $R_{AP}$ and $R_P$ denote the resistances in the anti-parallel and parallel magnetization configuration respectively. In Figure 4b, while the reading bias is -100mV, the TMR ratios are ~47 % ($TMR_{up}$) and ~12 % ($TMR_{down}$) for the upward and downward ferroelectric polarizations of PZT, respectively. The TER ratio is ~110 %. The interfacial spin polarization can be modified by the reversible ferroelectric polarization of PZT, which is the origin of the different TMR ratios.[8,43,44] The tunneling electromagnetoresistance (TEMR) indicating interfacial magnetoelectric coupling,[46,47] which is defined as ($TMR_{up}$ - $TMR_{down}$)/ $TMR_{down}$, is about 292 %. As a comparison, a typical symmetric MFTJ of LSMO/PZT/LSMO (or SPS, for short) has been fabricated and studied. The four distinct resistive states are also realized with the different ferroelectric polarizations of PZT (see Supplementary Figure 8b). The I-V behaviors and tunneling conductance with different ferroelectric polarizations at 60 K show no visible asymmetry (see Supplementary Figure 8a and 8c). Obviously, the introduction of the n-type LTMO electrode and LTMO/PZT interface play a critical role in the specific performance of the asymmetric SPT MFTJ.

The intrinsic diode rectification in the SPT MFTJ device provides an approach to generate approximate unidirectional conductivity (with different resistive states: "ON" and "OFF") as shown in Figure 4c. The writing and retention performances of the SPT device have also been confirmed in the absence of magnetic field at 60 K (see Figure 4d). Interestingly, with the coupling between TMR and TER, two different groups of four resistive states under opposite reading bias voltages (±100 mV) are realized in our device at 60 K (see Figure 4b). Reading with a positive bias, the TER ratio is 70 % while the TMR ratios are 21 % and 6 % for the upward and downward ferroelectric polarizations, respectively. Hence, the TEMR ratio with the positive reading bias of +100 mV is about 250 %. Compared with the TMR, TER and TEMR ratios with negative reading bias, all the reduced parameters with the positive one indicates that the spin-dependent tunneling and interfacial magnetoelectric coupling are weakened due to the electric field-driven diffusion and injection of the carriers from the manganate electrodes overcoming the built-in field.

**Device illustrations.** The unidirectional conductivity due to the intrinsic diode effect and nonvolatile multistate modulated by ferroelectric polarization and the reading bias can inspire us to design new kinds of array devices with the coexistence of logic units (AND and OR) and quaternary memory cells, as shown in Figure 5a. The logic cells consisting of two parallel asymmetric junctions with different connection modes and the appropriate series resistance $R_{OFF}$ are able to realize the AND and OR logical functions (the logic truth tables are shown in Figure 5b). Unlike the ferroelectric diodes, our asymmetric junctions can work without pre-writing to achieve ON and OFF resistive states, which brings about lower energy consumption and simpler circuit structure. Meanwhile, every asymmetric junction as a memory cell can store a quaternary message, and two group of quaternary bits can be stored and read with different bias polarity (see Figure 5b). Considering the independent physical origins of the unidirectional conductivity (due to p-i-n diode effect) and the multiple resistive states (from TMR and TER), the stored information will not be overwritten when the logical operation is performed under the appropriate bias voltages. The coexistence of logic units and quaternary memory cells in the same array devices can help to significantly increase the integration level of spintronic devices.

**Discussion**

To further understand the different tunneling effects (including TMR and TER) under opposite reading bias, some possible conduction mechanisms (for example, space-charge limited current (SCLC) and Ohmic conduction)[48-50] and Brinkman tunneling model[51] have been used to fit the I-V curves and tunneling conductance. Taking the fitting results under $P_{up}$ state for example, as shown in Figure 6, the conducting behavior during the whole negative bias region and low positive bias region (< 100 mV) can fit well with the SCLC mechanism ($I \propto V^2$) due to the interfacial space charge region and built-in field.[48] The current behavior in high positive bias region (> 100 mV) is associated with the Ohmic conduction ($I \propto V$) due to the diffusion and injection of major carriers,[49,52] which can weaken the spin-dependent tunneling and interfacial magnetoelectric coupling. The conducting behavior under $P_{down}$ state exhibits similar fitting results (see Supplementary Figure 9). As shown in Figure 6d, the asymmetry of normalized tunneling conductance $G(V)/G(0)$ for $P_{up}$ and $P_{down}$ states is so remarkable that only double Brinkman parabolic curves corresponding to two different groups of four resistive states can fit well with the experimental results. The fitting curves in negative bias region and low positive bias region (< 100 mV) demonstrate that the tunneling effect (or combined with SCLC mechanism) dominates the transport behavior in the device. The departure between the experiment and fitting curves in high positive bias region (> 100 mV) indicates the weakened tunneling effect. The conducting behavior around the critical positive bias (about 100 mV) has a complicated mechanism including SCLC, Ohmic conduction and tunneling effect. Referring to the Brinkman model, the different parabolic curves in negative and positive bias regions indicate the modified asymmetric interfacial barrier heights due to different polarities of the applied electric field which bring about the changes of the charge screening and potential profile (see Supplementary Figure 10).

We have shown that the asymmetric SPT MFTJ with the p-type and n-type electrodes exhibits the intrinsic diode rectification which is coexisting and coupling with the TMR and TER effects. Using the opposite reading bias polarities, two different groups of four resistive states have been realized. The spin-dependent tunneling and interfacial magnetoelectric coupling can be modulated by the ferroelectric polarization of PZT and the reading bias. Tunneling effect (or combined with SCLC) may dominate the transport behavior in negative bias and low positive bias regions (< 100 mV). The conducting behavior around the critical positive bias (about 100 mV) has a complicated mechanism including SCLC, Ohmic conduction and tunneling effect. The intrinsic diode effect combined with the TMR and TER effects will be of great significance for developing new types of spintronic devices with multi-states storage as well as logical operations.

**Methods**

**Sample growth.** The perovskite multilayers LSMO/PZT/LTMO were deposited in sequence on (100)-oriented $SrTiO_3$ (STO) substrates by pulsed laser deposition (PLD) using a KrF excimer laser (λ=248 nm) without reflection high-energy electron diffraction (RHEED). The commercial STO single crystal substrates with an atomically flat, single $TiO_2$-terminated surface were provided after preprocessing.

For the bottom LSMO layer, the growth conditions were maintained with 780 ºC substrate temperature, 10 Pa oxygen pressure, 1.14 J/cm$^2$ laser energy density and 3 Hz pulse repetition rate. A PZT ferroelectric barrier was then grown at 700 ºC substrate temperature, 10 Pa oxygen pressure, 1 J/cm$^2$ laser energy density and 3 Hz pulse repetition rate. Finally, a top LTMO layer was deposited with the conditions of 725 ºC substrate temperature, 10 Pa oxygen pressure, 1.14 J/cm$^2$ laser energy density and 3 Hz pulse repetition rate. After the deposition, the multilayers were *in situ* annealed at 700 ºC under an oxygen pressure of 10000 Pa for 10 min to reduce oxygen vacancies. MFTJ devices of a square shape (20 μm ×20 μm) were patterned by electron beam lithography (EBL), ion-milling etching and lift-off process. A polycrystalline Al$_2$O$_3$ film used to isolate the bottom LSMO electrode was deposited by RF sputtering, and the Pt electrodes for bonding were deposited by DC sputtering.

**Magnetic and electric measurements.** The magnetic property of multilayers was measured by applying an in-plane magnetic-field parallel to the substrate [110] direction at 60 K using a vibrating sample magnetometer (VSM) option in a versalab system. The R vs H behaviors were carried out with an in-plane magnetic-field parallel to the substrate [110] direction at 60 K using Keithley 2400 connected with a versalab system. The pulsed writing bias (5 V) and reading bias (100 mV) are of the same pulse width 100 ms. Using Keithley 6221 and 2182A, the I-V curves and conductance of MFTJs were measured in the voltage sweep mode with the measurement integration of 4, repeat filter count of 2 and delay time of 2 ms.

**Electrical analysis.** Some possible conduction mechanisms have been used to fit the I-V curves.[48-49] Ohmic conduction showing a linear relationship between voltage and current is given by

$$J_{Ohm} = \sigma E \quad ,$$

where $\sigma$ is the electrical conductivity.[49] The space-charge limited current (SCLC) is described as the following equation.

$$J_{SCLC} = \frac{9\mu\varepsilon_r\varepsilon_0}{8d}E^2 \quad ,$$

where $\mu$ is drift mobility of charge carrier, $\varepsilon_r$ the permittivity of dielectric, and $\varepsilon_0$ the permittivity of free space.[48] The Brinkman model[51] used to analyze tunneling conductance is described by

$$G(V)/G(0) = \left(1 - \frac{d\sqrt{2m_e}}{12\hbar\sqrt{\bar{\phi}}}\frac{\phi}{\bar{\phi}}V + \frac{d^2 m_e e}{4\hbar\bar{\phi}}V^2\right) \quad ,$$

where $G(0)$ is the tunneling conductance at zero bias, $d$ the effective barrier thickness, $m_e$ the effective mass, $\phi$ the difference of the asymmetric interfacial barrier height and $\bar{\phi}$ the average barrier height.

**Data availability**

The data that support the findings of this study are available from the corresponding authors upon request.

**Figure 1 | Structural and ferroelectric characterizations.** (**a**) AFM image of a fabricated multilayer of LSMO (30 nm)/PZT (3.6 nm)/LTMO (15 nm). (**b**) PFM image of a LSMO(30 nm)/PZT(3.6 nm) bilayer after being written under ± 5 V. (**c**) Ferroelectric phase hysteresis and strain loops of the LSMO(30 nm)/PZT(3.6 nm) bilayer. (**d**) RSM image of the multilayer of LSMO/PZT/LTMO around (103) reflection.

**Figure 2 | Interfacial structure and chemical mapping.** (**a**) HRTEM-HAADF micrograph of the interfaces in LSMO/PZT/LTMO, with the inset revealing the displacement of a Ti ion in a PZT unit cell, confirming the upward spontaneous polarization. The green and red spheres denote Ti and Pb ions, respectively. (**b**) Elemental profiles across the yellow box area of (**a**) obtained from the EELS results. (**c,d**) Power-law subtracted EELS spectra of the Mn $L_{2,3}$ edge corresponding to the interfacial manganite layers: (**c**) top LTMO and (**d**) bottom LSMO. The red arrows and layer numbers indicate the scan direction from the PZT interface into the manganite layers corresponding to the arrows and numbers of (**a**).

**Figure 3 | EELS analysis of the LSMO/PZT/LTMO structure.** (**a,b**)The peak separation ($\Delta E_{\text{Mn } L2,3}$) between Mn $L_{2,3}$ edge peak positions (blue solid squares) and $L_3/L_2$ peak intensity ratio (rose hollow squares) for the top LTMO (**a**) and bottom LSMO (**b**) layers near the PZT interfaces, respectively. (**c,d**) Mapping of Mn valence (red solid spheres) in the vicinity of the PZT interfaces for the (**c**) top LTMO and (**d**) bottom LSMO, respectively.

**Figure 4 | Electrical characterization I of the SPT MFTJ under 60 K.** (**a**) I-V curves under different polarizations. The inset shows the schematic of the measurement. (**b**) R-V curves under different polarizations: upward (blue) and downward (red) obtained from (**a**). The inset shows the enlarged curves near 0.1 V. (**c**) Repeatable four resistive states using different writing and reading bias voltages in the absence of magnetic field. (**d**) TMR curves under different polarizations and different

reading bias: upward polarization (blue), downward polarization (red), negative bias (solid) and positive bias (hollow). The insets show the schematics of the p-i-n diode. The horizontal black arrows indicate the magnetization directions of two electrodes.

**Figure 5 | Device illustrations.** (**a**) The blueprint of the array devices with the coexistence of logic units (AND and OR gates) and quaternary memory cells. (**b**)The logic truth tables of AND and OR gates and the schematics for the two groups of quaternary bits.

**Figure 6 | Electrical analysis of the SPT MFTJ in $P_{up}$ state.** I vs $V^2$ behavior and the fitting line in different bias regions obtained from Figure 4a: (**a**) negative bias region and (**b**) positive bias region I (< 0.1 V). (**c**) I vs V behavior and the fitting line in positive bias region II (>0.1 V). (**d**) Normalized conductance under different polarizations: upward (blue) and downward (red). The dashed parabolic curves indicate the fitting results adapted to Brinkman model.

**Additional Information**

Supplementary Information is available for this manuscript.

**Competing Interests**

The authors declare no competing financial interest.

**Correspondences**

Correspondence should be addressed to: Prof. Jun Miao (j.miao@ustb.edu.cn), Prof. Yong Jiang (yjiang@ustb.edu.cn).

**Author Contributions**

J. M. and Y. J. proposed the idea. J. M., Y. J. and Q. L. designed the experiments and wrote the manuscript. Q. L. performed film growth, and prepared the devices. Q. L. measured AFM and PFM assisted by P. F. L. Z. D.X. measured RSM and assisted to develope the film deposition strategy. Q. L. and P. F. L. managed the synchrotron XAS experiments. Q. H. Z and L.G measured TEM and EELS. Q. L and K. K .M characterized the transportation performance. J. M., Y. J. and Q. L. contributed to the data analysis. J. K. C, X. G. X and Y. W provided experimental support and useful discussions.

**Acknowledgements**

This work was partially supported by the National Basic Research Program of China (2015CB921502), the National Science Foundation of China (Grant Nos. 51731003, 11574027, 51671019, 51602022, 61674013, 51602025).

**Figure 1**

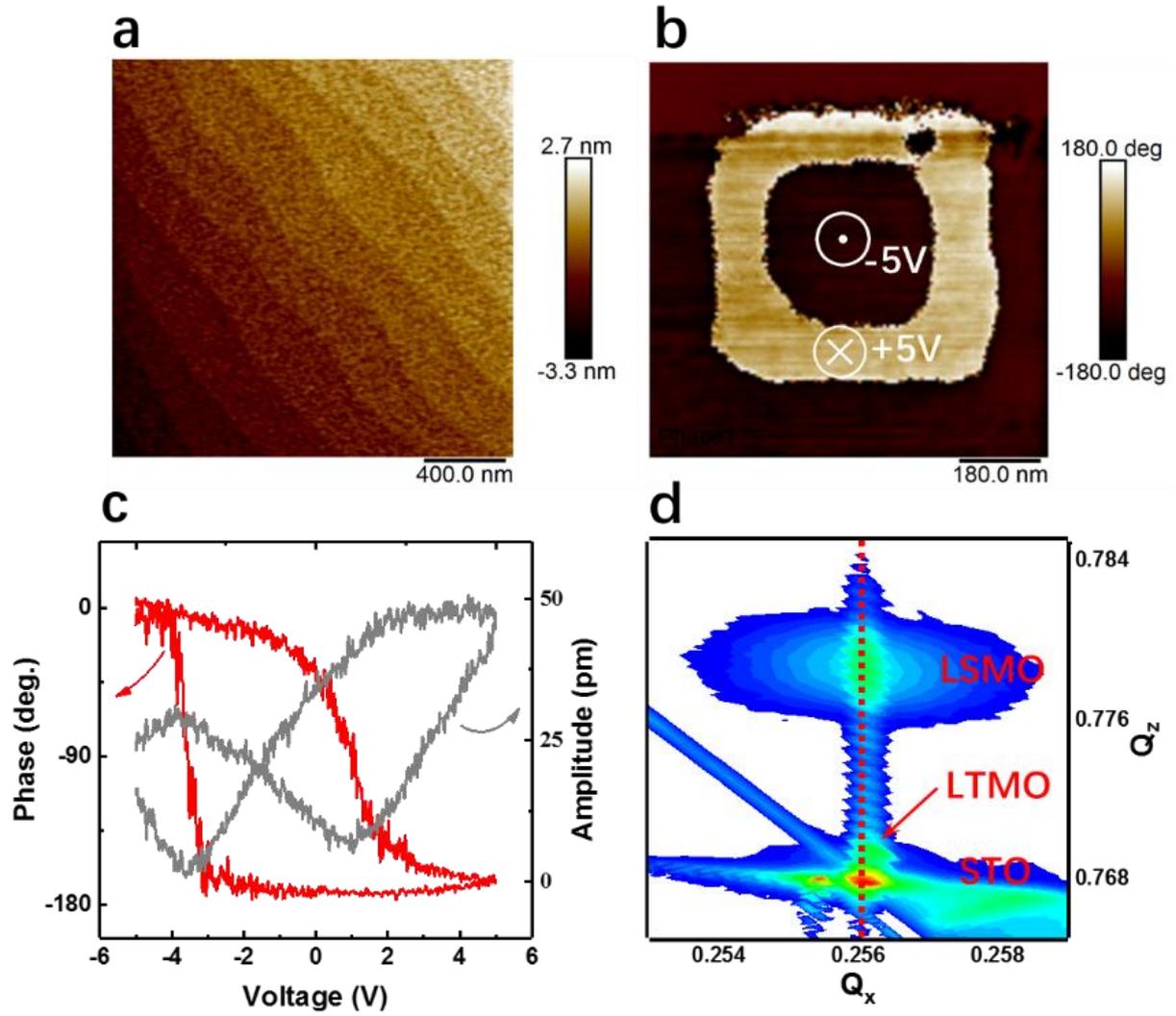

**Figure 2**

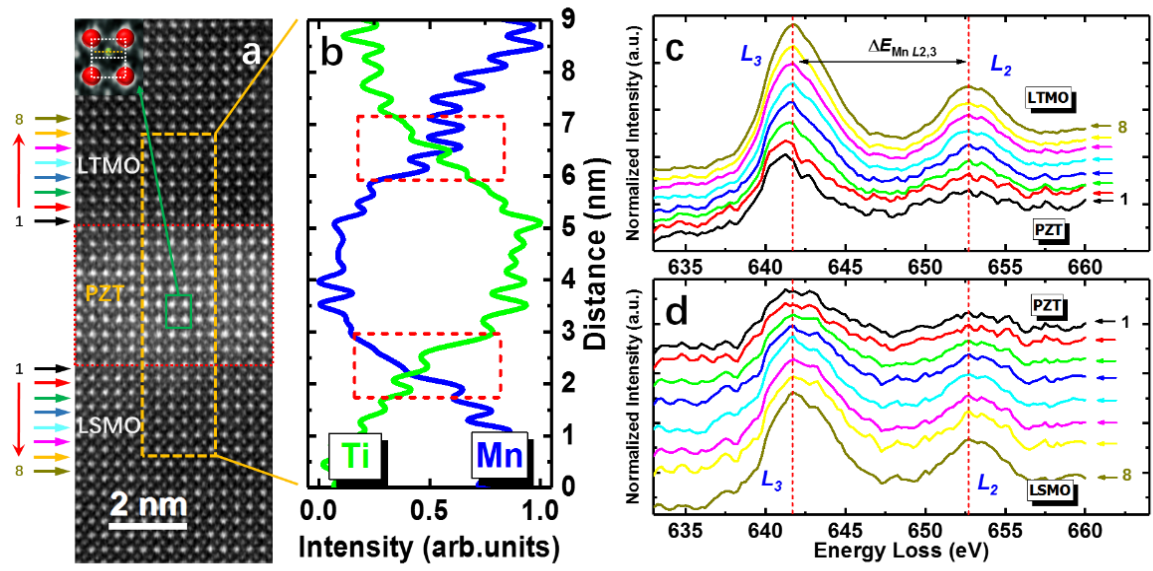

**Figure 3**

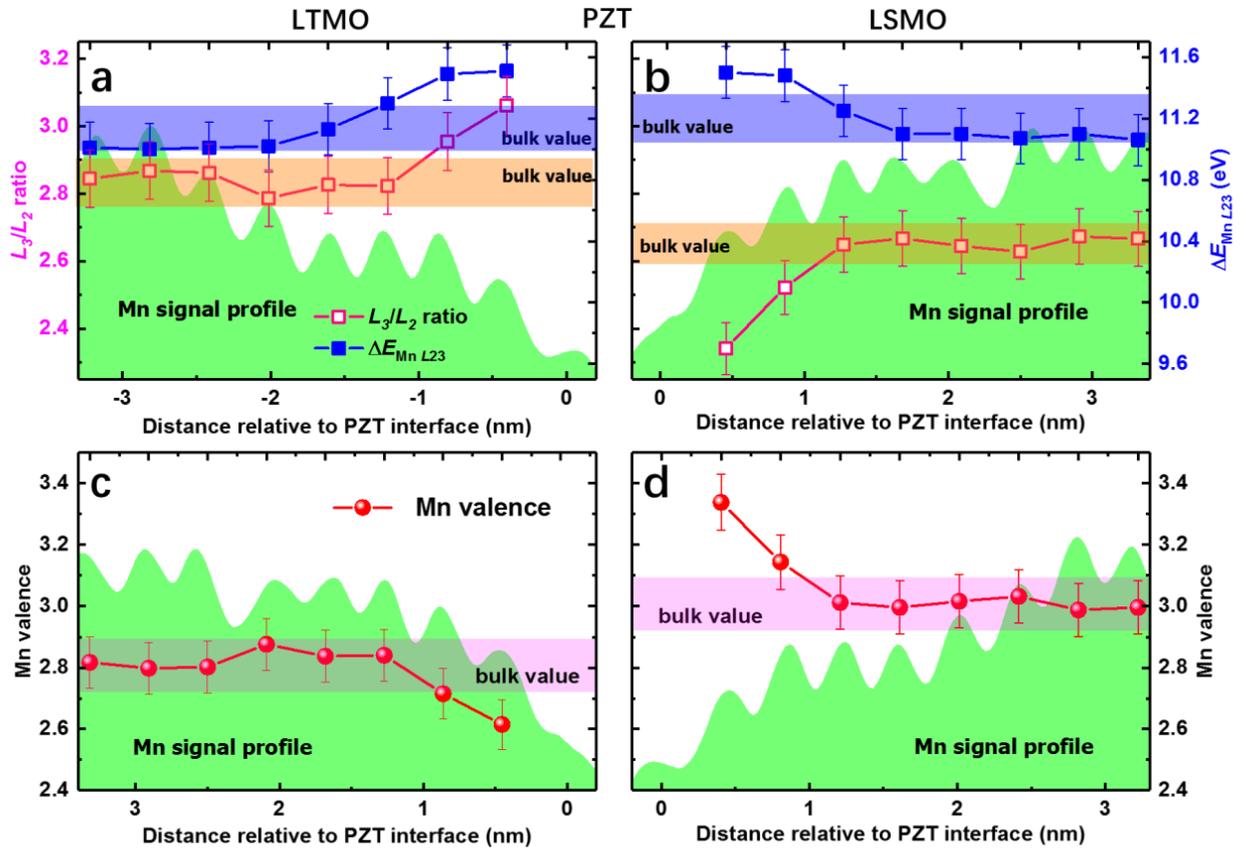

**Figure 4**

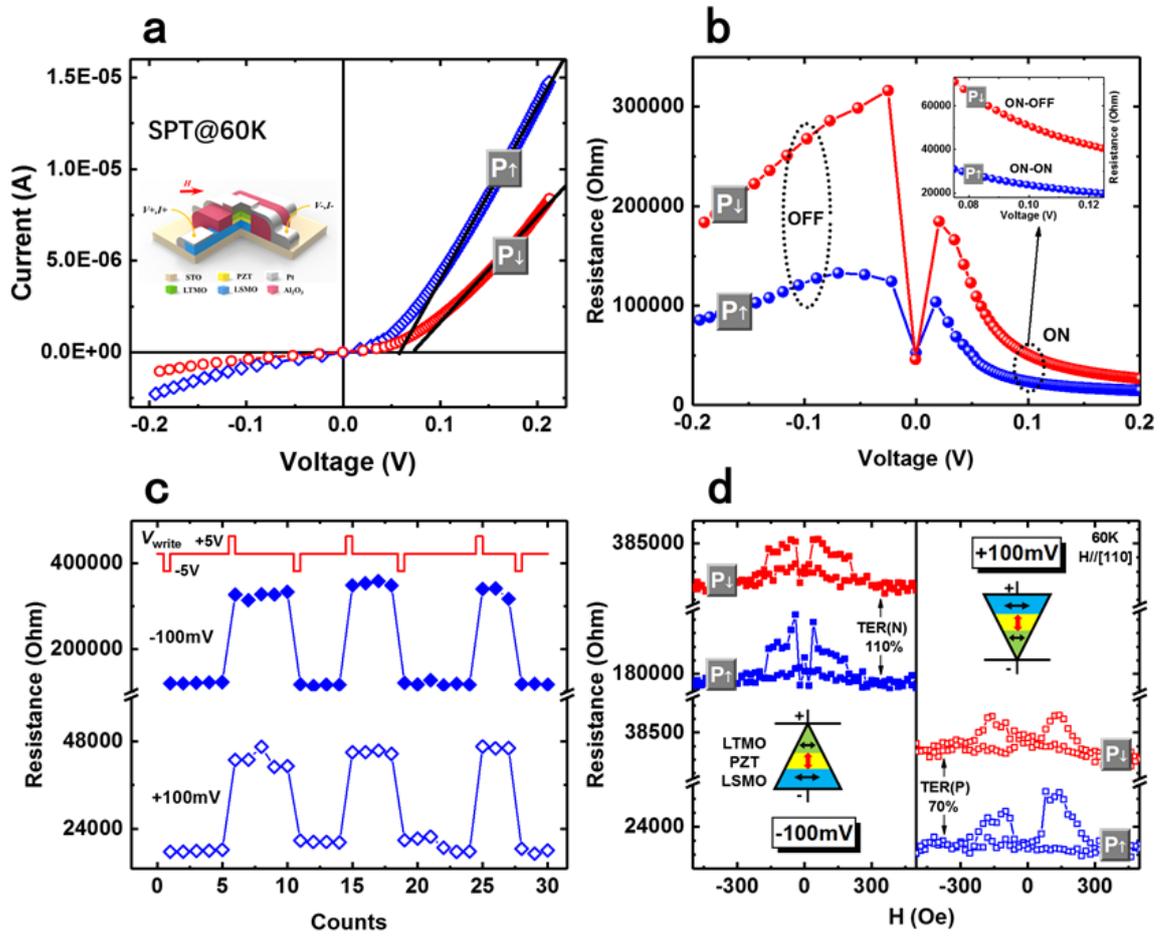

**Figure 5**

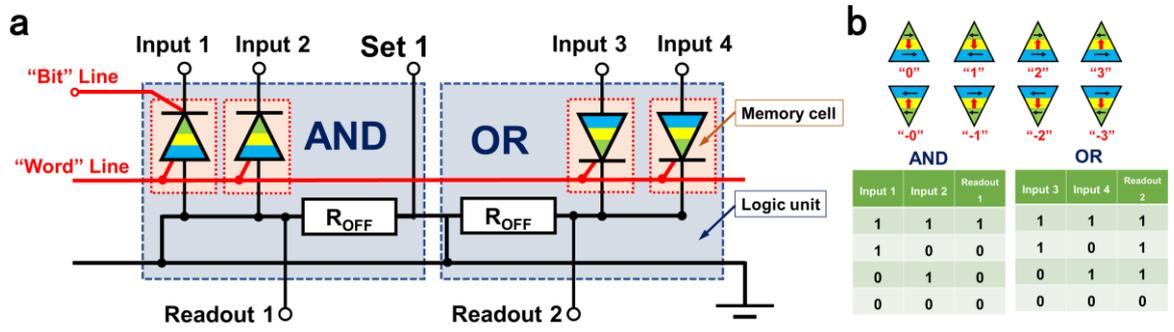

**Figure 6**

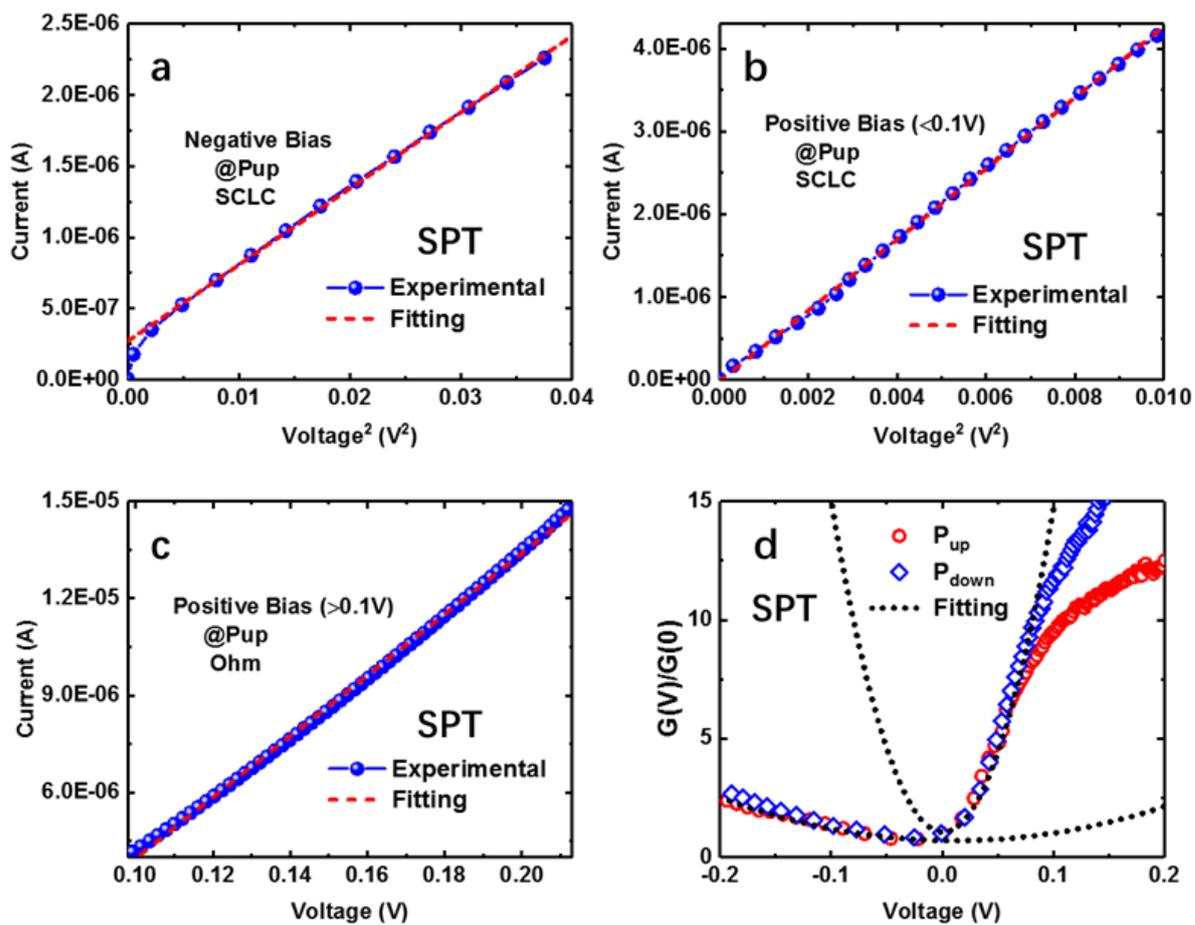